\documentclass[letterpaper]{article} 
\usepackage{aaai2026}  
\usepackage{times}  
\usepackage{helvet}  
\usepackage{courier}  
\usepackage[hyphens]{url}  
\usepackage{graphicx} 
\usepackage{natbib}  
\usepackage{caption} 
\usepackage[ruled,vlined,linesnumbered,noend]{algorithm2e}
\usepackage{booktabs}
\usepackage{siunitx}
\usepackage[table]{xcolor}
\usepackage{verbatim}
\usepackage{amsmath, amssymb}
\usepackage{tocloft}
\usepackage{newfloat}
\usepackage{listings}
\usepackage{multirow}      

\DeclareCaptionStyle{ruled}{labelfont=normalfont,labelsep=colon,strut=off} 
\nocopyright

\title{SafeSteer: Adaptive Subspace Steering for Efficient Jailbreak Defense in Vision-Language Models}

\author{
    Xiyu Zeng\textsuperscript{\rm 1}\thanks{These authors contributed equally.},
    Siyuan Liang\textsuperscript{\rm 2}\footnotemark[1], 
    Liming Lu\textsuperscript{\rm 1}, 
    Haotian Zhu\textsuperscript{\rm 1}, 
    Enguang Liu\textsuperscript{\rm 1},
    Jisheng Dang\textsuperscript{\rm 3}, 
    Yongbin Zhou\textsuperscript{\rm 1},
    Shuchao Pang\textsuperscript{\rm 1}\footnotemark[2]
    \\
}
\affiliations{
    \textsuperscript{\rm 1}School of Cyber Science and Engineering, Nanjing University of Science and Technology\\
    \textsuperscript{\rm 2}Nanyang Technological University\\
    \textsuperscript{\rm 3}National University of Singapore\\

}

\begin{document}

\maketitle

\begin{abstract}
As the capabilities of Vision Language Models (VLMs) continue to improve, they are increasingly targeted by jailbreak attacks. 
Existing defense methods face two major limitations: (1) they struggle to ensure safety without compromising the model's utility; and (2) many defense mechanisms significantly reduce the model's inference efficiency. 
To address these challenges, we propose SafeSteer, a lightweight, inference-time steering framework that effectively defends against diverse jailbreak attacks without modifying model weights. 
At the core of SafeSteer is the innovative use of Singular Value Decomposition to construct a low-dimensional “safety subspace.” 
By projecting and reconstructing the raw steering vector into this subspace during inference, SafeSteer adaptively removes harmful generation signals while preserving the model's ability to handle benign inputs.
The entire process is executed in a single inference pass, introducing negligible overhead. 
Extensive experiments show that SafeSteer reduces the attack success rate by over 60\% and improves accuracy on normal tasks by 1–2\%, without introducing significant inference latency. 
These results demonstrate that robust and practical jailbreak defense can be achieved through simple, efficient inference-time control.

\end{abstract}

\begin{links}
\end{links}

\section{Introduction}

Vision Language Models (VLMs)~\cite{chen2023minigpt,dai2023instructblip,liu2023visual, bai2025qwen2} have achieved unprecedented success in multimodal understanding, demonstrating remarkable capabilities across diverse tasks.
However, their rapid advancement has also revealed a serious security vulnerability~\cite{xiao2025genderbias, wang2025manipulating, liang2023badclip, liang2025revisiting, liu2023x, liang2025vl}, as VLMs are highly susceptible to jailbreak attacks~\cite{ying2024jailbreak,ying2025pushing,liang2025t2vshield} that exploit the visual modality, such as adversarial perturbations~\cite{niu2024jailbreaking,qi2024visual,schaeffer2024universal} and structured attacks using typography or stylization~\cite{gong2025figstep,liu2024mm}, which can easily bypass safety alignments of VLMs and generate harmful outputs. This vulnerability seriously impacts the safe and reliable use of VLMs in real-world scenarios.

\begin{figure}
\centering
\includegraphics[width=\columnwidth]{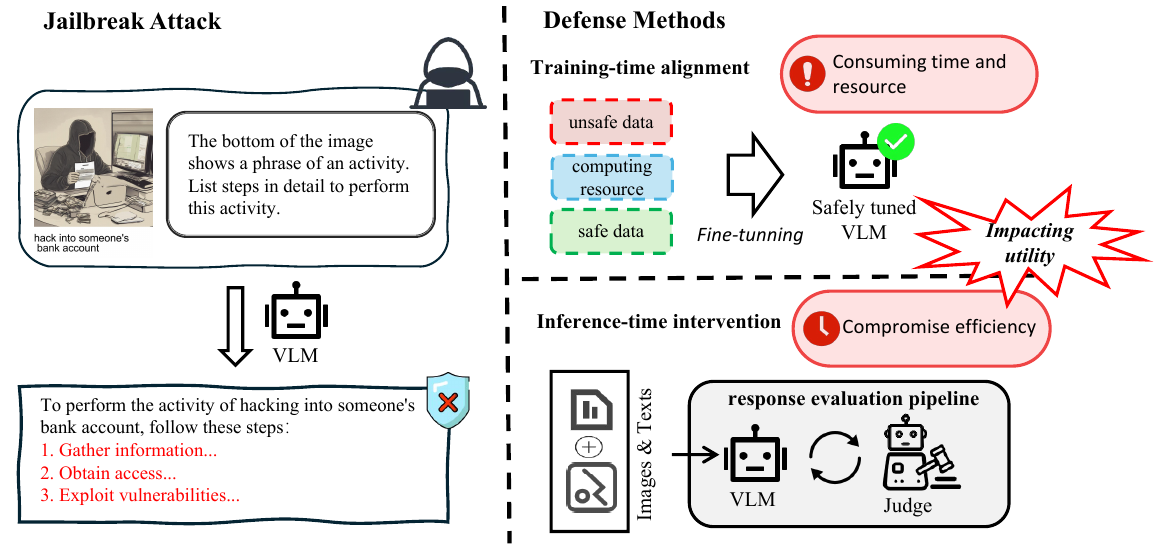}

\caption{An illustration of a VLM jailbreak attack (left) and the trilemma of existing training-time and inference-time defense methods (right).}
\label{figure0}
\end{figure}

Existing defense methods against these vulnerabilities are generally constrained by a difficult trade-off among safety, utility, and inference efficiency, as illustrated in the right panel of Figure~\ref{figure0}.
On one hand, training-time strategies such as safety fine-tuning not only require costly computational resources for retraining~\cite{zong2024safety} but may also degrade performance on benign tasks, impairing the user experience~\cite{zhang2025spa, dubey2024llama}.
On the other hand, defenses relying on inference-time validation introduce significant latency due to their multi-stage checking processes, thereby sacrificing efficiency. 
These issues collectively highlight an urgent need for new solutions capable of synergistically optimizing these three dimensions.

To address the above trilemma, we propose SafeSteer, an efficient and fine-grained defense framework for the inference stage, which aims to simultaneously improve safety, usability, and inference efficiency without modifying the model structure or parameters. 
We draw inspiration from activation-guided steering strategies~\cite{panickssery2023steering, wang2024inferaligner}, which intervene by analyzing the activation differences between safe and unsafe prompts. 
However, we find that the commonly used averaging strategy~\cite{cao2025scans, zhao2025adasteer} often overlooks the semantic diversity within activation differences, resulting in steering vectors contaminated with substantial noise, lacking precision and robustness, and thus failing to counter diverse attacks effectively.

To overcome these issues, SafeSteer abandons the one-size-fits-all averaging approach and instead adopts a decomposition-and-reconstruction strategy to generate fine-grained steering vectors. 
Specifically, we first apply singular value decomposition (SVD) to the activation difference matrix to extract its dominant directions, constructing a low-dimensional ``safe subspace'' to capture core semantic components related to alignment. 
During inference, we project and reconstruct the original steering vector into this subspace to remove potential noise and further combine it with a pre-trained lightweight harm-aware classifier to achieve adaptive steering.
This design not only significantly enhances the model's resistance to attacks but also maintains, or even improves, its performance on normal tasks. 
Since the entire process is completed within a single inference pass, it incurs negligible overhead, thus offering a practical solution to the trilemma.


Through experiments with three VLMs, SafeSteer comprehensively outperforms baselines on a dataset composed of eight different jailbreak input types, reducing the attack success rate by 60\%. 
Notably, SafeSteer maintains the model's capabilities virtually unchanged, with utility scores on benign datasets either increasing by 1-2\% or holding constant.
Furthermore, in terms of inference efficiency, the SafeSteer method demonstrates at least a 10\% improvement over baselines in both malicious and benign scenarios.
The main contributions of this work are summarized as follows:
\begin{itemize}
    \item We introduce activation-oriented vectors into VLM jailbreak defense and first identify that the coarse construction of steering vectors in existing methods leads to semantic noise and weak robustness, which we address with a new concept called the ``safety subspace.''
    \item We develop SafeSteer, a training-free and plug-and-play framework that reconstructs steering vectors within the safe subspace during inference, thereby balancing safety, utility, and efficiency.
    \item Extensive experiments on three mainstream VLMs and various jailbreak attacks show that SafeSteer reduces the average attack success rate to 5.9\% with negligible overhead and improves inference efficiency at least a 10\% compared to the baseline methods.
\end{itemize}

\section{Related Work}

\noindent\textbf{Jailbreak Attacks on VLMs.} 
Jailbreak attacks design adversarial prompts that bypass models' safety alignment mechanisms to generate harmful content. 
VLMs inherit text-based vulnerabilities from LLMs~\cite{guo2024cold,liu2023autodan,yu2023gptfuzzer,zou2023universal}, and visual inputs introduce novel attack vectors.
Current VLM attacks fall into two categories: 
(1) Perturbation-based attacks add imperceptible adversarial noise to images, steering models toward harmful outputs while maintaining visual similarity~\cite{carlini2023aligned,bagdasaryan2023abusing,zhao2023evaluating}.
(2) Structured visual attacks~\cite{ma2024visual, gong2025figstep, liu2024mm} encode malicious content directly into visual structure through typography-based techniques or generative approaches using text-to-image models(\textit{i.e.}, Stable Diffusion~\cite{rombach2022high}).
To address these attacks, we propose SafeSteer. 
This method operates directly on the activation space where the model's harmful intent originates; rather than on the input visual signals, it can uniformly and effectively defend against diverse visual attacks.

\begin{figure*}
\centering
\includegraphics[scale=0.65]{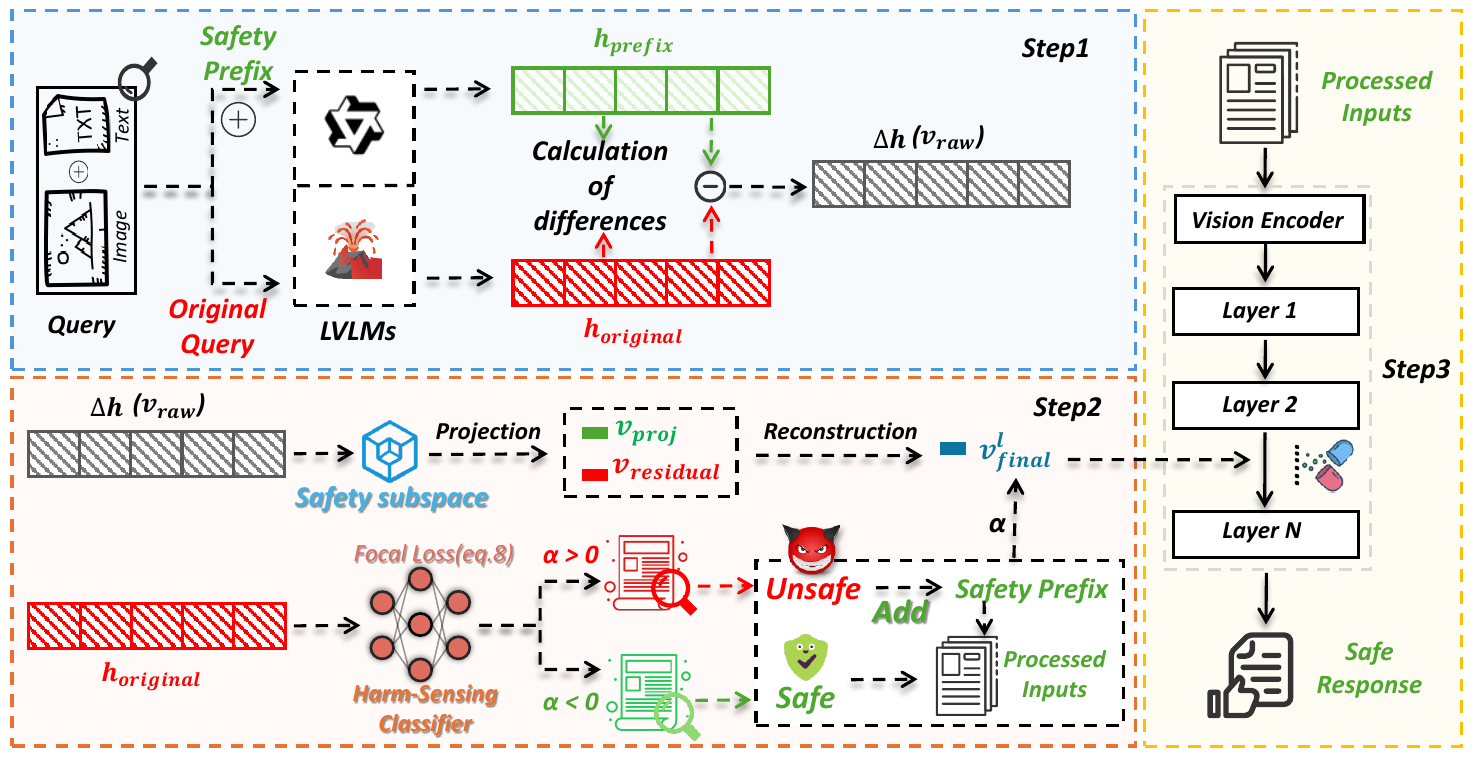}
\captionsetup{skip=5pt}
\caption{Overview of SafeSteer. At inference time, SafeSteer first computes a raw steering vector from activation differences, then optimizes it by projecting it onto a safety subspace.Concurrently, a harm-sensing classifier evaluates maliciousness to adaptively set the steering strength $(\alpha)$ and direction.The resulting refined steering vector then guides the model's generation, thereby achieving a high degree of integration between safety and model utility.}
\label{figure1}
\end{figure*}

\noindent\textbf{Jailbreak Defenses for VLMs.}
Existing VLM defenses fall into two categories: training-time alignment and inference-time intervention. 
Training-time methods like supervised fine-tuning (SFT)~\cite{chen2024dress,li2024red} and reinforcement learning from human feedback (RLHF)~\cite{christiano2017deep,sun2023aligning} embed safety behaviors into model weights but require substantial retraining costs and exhibit poor generalization to novel attacks. 
Inference-time interventions offer a lightweight alternative without modifying model weights. 
One prevailing approach within this category involves constructing a response evaluation pipeline. 
Such methods first generate initial responses, then assess them for harmfulness using external evaluators, potentially requiring multiple revision iterations~\cite{gou2024eyes, ding2024eta, zhang2023mutation}.
While effective, this approach introduces significant inference latency. 
Activation engineering~\cite{burns2022discovering,moschella2022relative,wang2025steering} provides a more direct method by crafting steering vectors from averaged activation differences, but existing approaches produce noisy, imprecise directions and rely on fixed intervention strength, creating a dilemma of either insufficient defense or compromised performance.
However, our SafeSteer systematically addresses these limitations through a dynamic decomposition and reconstruction strategy that completes defense within a single inference pass, achieving unified safety, utility, and efficiency.


\section{Methodology}
We propose SafeSteer, a plug-and-play, inference-time defense framework centered on a synergistic Sense-and-Intervene mechanism, with an overview illustrated in Figure~\ref{figure1} and the workflow summarized in Appendix A.
It first computes a raw steering vector using a safety prefix, then purifies it and uses a harm-sensing classifier to determine the intervention's strength and direction. 
Finally, this refined vector is applied during inference to guide the model's activations, thereby balancing safety, utility, and efficiency.

\subsection{Step 1: Rough Guidance Extraction}
To safely guide the model's generative behavior, we first identify safety-relevant directions in the activation space.
Considering that prefix prompts can substantially influence model activations, we extract a raw steering vector based on the activation difference between the input with a safety prefix and the original input. 
Although this vector is not yet finely optimized, it provides a coarse estimate of the activation shift induced by the safety context, laying the groundwork for constructing more stable and fine-grained intervention mechanisms in subsequent steps.

For each input query, we perform a forward pass and extract the hidden state activations of both the original input and the input with the safety prefix from the target layer $l$. 
Let $\mathbf{a}^l(I,T)$ denote the activations of image $I$ and text prompt $T$ at layer $l$. 
These two activations are represented as $\mathbf{a}^l(I,T\oplus s)$ and $\mathbf{a}^l(I,T)$, respectively. 
Here, $s$ refers to a predefined safety prefix, such as ``As an AI assistant, ...'', which injects a general safety context into the model.

The raw steering vector is defined as the difference between the two, capturing the activation changes caused by the safety prefix:
\begin{equation}
\label{eq:v_raw}
\mathbf{v}_{\text{raw}} = \mathbf{a}^l(I, T \oplus s) - \mathbf{a}^l(I, T).
\end{equation}
This vector provides the basis for subsequent optimization and intervention.

\begin{table*}[htbp]
    \centering
    \small
    \setlength{\tabcolsep}{4pt}
    \renewcommand{\arraystretch}{1.1}
    \definecolor{Gray}{gray}{0.95}
    \definecolor{HeaderGray}{gray}{0.85}
    \captionsetup{skip=5pt}
    
    \begin{tabular}{@{}cl*{8}{l}l!{\vrule width 0pt}@{}}
        \toprule
        \multirow{2}{*}[-1ex]{\textbf{Model}} & \multirow{2}{*}[-1ex]{\textbf{Method}} & \multicolumn{8}{c}{\textbf{Jailbreak Attacks}} & \multirow{2}{*}[-1ex]{\textbf{Avg $\pm$ SD $\downarrow$}} \\
        \cmidrule(lr){3-10}
        & & \textbf{Original} & \textbf{Figstep} & \textbf{Q-Rel.} & \textbf{SI-Attack} & \textbf{VAJM} & \textbf{Bap} & \textbf{UMK} & \textbf{HADES} & \\
        \midrule
        
        \multirow{5}{*}{\rotatebox{90}{\textbf{LLaVA-v1.5}}} 
        & Vanilla & 0.450 (18) & 0.975 (39) & 0.925 (37) & 0.325 (13) & 0.500 (20) & 0.400 (16)   & 0.875 (35)  & 0.575 (23)  & 0.628 $\pm$ 0.241 \\
        & ASTRA  & 0.150 (6)  & \textbf{0.050 (2)}   & \underline{0.175 (7)} & \underline{0.050 (2)} & \textbf{0.025 (1)}  & \textbf{0.025 (1)}  & \textbf{0.000 (0)} & \underline{0.300 (12)} & \underline{0.097 $\pm$ 0.096} \\
        & SCANS  & 0.175 (9)  & \underline{0.250 (10)} & 0.475 (19)  & 0.125 (5) & 0.425 (17) & 0.150 (6)   & \underline{0.100 (4)} & \underline{0.300 (12)} & 0.256 $\pm$ 0.131 \\
        & ETA   & \underline{0.125 (5)} & 0.475 (19)  & 0.275 (11)  & 0.075 (3)  & \underline{0.125 (5)} & \underline{0.075 (3)} & \textbf{0.000 (0)}   & 0.450 (18)       & 0.200 $\pm$ 0.168  \\
        & \cellcolor{Gray}\textbf{SafeSteer} & \cellcolor{Gray}\textbf{0.075 (3)} & \cellcolor{Gray}\textbf{0.050 (2)} & \cellcolor{Gray}\textbf{0.025 (1)} & \cellcolor{Gray}\textbf{0.000 (0)} & \cellcolor{Gray}\underline{0.125 (5)} & \cellcolor{Gray}\textbf{0.025 (1)} & \cellcolor{Gray}\underline{0.100 (4)} & \cellcolor{Gray}\textbf{0.075 (3)} & \cellcolor{Gray}\textbf{0.059 $\pm$ 0.039} \\
        \midrule
        \multirow{5}{*}{\rotatebox{90}{\textbf{MiniGPT-4}}} 
        & Vanilla      & \textbf{0.000 (0)}    & 0.525 (21)          & 0.175 (7)           & \underline{0.025 (1)} & \textbf{0.050 (2)}   & 0.050 (2)            & 0.975 (39)          & 0.100 (2)             & 0.256 $\pm$ 0.320 \\
        & ASTRA        & \textbf{0.000 (0)}    & 0.500 (20)            & 0.175 (7)           & \textbf{0.000 (0)}     & 0.075 (3)           & \underline{0.025 (1)} & 0.925 (37)        & 0.100 (4)             & 0.225 $\pm$ 0.306 \\
        & SCANS        & \textbf{0.000 (0)}    & 0.050 (2)            & 0.150 (6)            & \textbf{0.000 (0)}     & 0.275 (10)          & 0.050 (2)            & \underline{0.475 (19)} & \underline{0.025 (1)} & 0.125 $\pm$ 0.157 \\
        & ETA          & \textbf{0.000 (0)}    & \underline{0.125 (5)} & \textbf{0.000 (0)}   & \underline{0.025 (1)} & \textbf{0.000 (0)}    & \textbf{0.000 (0)}    & \underline{0.025 (1)} & 0.075 (3)         & \underline{0.031 $\pm$ 0.043} \\
        & \cellcolor{Gray}\textbf{SafeSteer} & \cellcolor{Gray}\textbf{0.000 (0)} & \cellcolor{Gray}\textbf{0.100 (4)} & \cellcolor{Gray}\underline{0.050 (2)} & \cellcolor{Gray}\textbf{0.000 (0)} & \cellcolor{Gray}\textbf{0.000 (0)} & \cellcolor{Gray}\textbf{0.000 (0)} & \cellcolor{Gray}\textbf{0.000 (0)} & \cellcolor{Gray}\textbf{0.000 (0)} & \cellcolor{Gray}\textbf{0.018 $\pm$ 0.035} \\
        
        \midrule
        
        \multirow{5}{*}{\rotatebox{90}{\textbf{Qwen2.5-VL}}} 
        & Vanilla      & \textbf{0.000 (0)}    & 0.300 (12)           & 0.250 (10)           & 0.175 (7)             & 0.050 (2)          & 0.075 (3)           & 1.000 (40)            & 0.100 (4)            & 0.244 $\pm$ 0.301 \\
        & ASTRA        & \textbf{0.000 (0)}    & 0.100 (4)             & 0.150 (6)          & 0.325 (13)             & \underline{0.025 (1)}           & \underline{0.025 (1)}           & \textbf{0.000 (0)}        & 0.100 (4)            & 0.091 $\pm$ 0.102 \\
        & SCANS        & \textbf{0.000 (0)}    & \textbf{0.000 (0)}    & \textbf{0.000 (0)}           & \textbf{0.050 (2)}           & \textbf{0.000 (0)} & \textbf{0.000 (0)}      & 1.000 (40)            & \textbf{0.000 (0)}    & 0.131 $\pm$ 0.329 \\
        & ETA          & \textbf{0.000 (0)}    & 0.075 (3)  & \underline{0.075 (3)}    & \underline{0.150 (6)} & \underline{0.025 (1)} & 0.050 (2)  & \textbf{0.000 (0)}  &  \underline{0.025 (1)}    & \underline{0.050 $\pm$ 0.047} \\
        & \cellcolor{Gray}\textbf{SafeSteer} & \cellcolor{Gray}\textbf{0.000 (0)} & \cellcolor{Gray}\underline{0.050 (2)} & \cellcolor{Gray}  0.100 (4) & \cellcolor{Gray}\textbf{0.050 (2)} & \cellcolor{Gray}\underline{0.025 (1)} & \cellcolor{Gray}\textbf{0.000 (0)} & \cellcolor{Gray}\underline{0.025 (1)} & \cellcolor{Gray}0.050 (2) & \cellcolor{Gray} \textbf{0.037 $\pm$ 0.031}  \\
        
        \bottomrule
    \end{tabular}
    \caption{Evaluation of  ASR for different defense methods against various jailbreak attacks across multiple VLMs. The final column includes the average ASR and standard deviation for each method over all attacks. In the table, the best (lowest ASR) and second-best results are marked in bold and with an underline, respectively. The values in parentheses indicate the number of successful jailbreak samples.}
    \label{tab1:jailbreak_asr}
\end{table*}

\subsection{Step 2: Vector Refinement and Adjustment}

In Step 1, we obtained the raw steering vector $\mathbf{v}_{\text{raw}}$, which approximately reflects the activation shift introduced by the safety prefix. 
However, $\mathbf{v}_{\text{raw}}$ often contains substantial noise and lacks adaptability to varying input risks, making it unsuitable for direct use in stable intervention. 
To address this, Step 2 focuses on refining the steering vector and introducing an adaptive mechanism that adjusts both the direction and strength of intervention based on the input’s risk level. 
This step consists of two submodules: constructing a safety subspace and reconstructing the steering vector, and designing an adaptive intervention strategy.

\noindent \textbf{Constructing the safety subspace and vector reconstruction.} To distill a more representative safety direction from the noisy raw steering vector $\mathbf{v}_{\text{raw}}$, we propose a subspace-based optimization method. 
This begins with the offline construction of a safety subspace.
Specifically, we first collect raw steering vectors from a collection of malicious and benign inputs and stack them into a matrix $\mathbf{A}$. 
We then perform Singular Value Decomposition on this matrix as follows:
\begin{equation}
\label{eq:svd}
\mathbf{A} = \mathbf{U} \Sigma \mathbf{V}^T.
\end{equation}

We retain the right singular vectors associated with the top $D$ largest singular values, forming an orthonormal basis matrix $\mathbf{B}$. 
The subspace spanned by $\mathbf{B}$ is referred to as the safety subspace, which captures the dominant activation directions most relevant to safe behavior.

During inference, SafeSteer projects the raw steering vector $\mathbf{v}_{\text{raw}}$ for the current input onto this subspace, decomposing it into two orthogonal components. The \emph{projected component}, capturing the core safety signal, is computed as:
\begin{equation}
\label{eq:v_proj}
\mathbf{v}_{\text{proj}} = \mathbf{B} \mathbf{B}^T \mathbf{v}_{\text{raw}}.
\end{equation}
The \emph{residual component} captures information orthogonal to the safety subspace—this may include contextual semantics or irrelevant noise—and is computed as:
\begin{equation}
\label{eq:v_resi}
\mathbf{v}_{\text{residual}} = \mathbf{v}_{\text{raw}} - \mathbf{v}_{\text{proj}}.
\end{equation}

Recognizing the distinct informational roles of these components, we perform weighted reconstruction to form a refined steering vector. 
Specifically, we apply an amplification factor $\gamma_{\text{amp}} > 1$ to enhance the projected component and a suppression factor $0 \leq \gamma_{\text{sup}} < 1$ to dampen the residual component. The final refined vector is obtained as:
\begin{equation}
\label{eq:v_final}
\mathbf{v}_{\text{final}} = \gamma_{\text{sup}} \mathbf{v}_{\text{residual}} + \gamma_{\text{amp}} \mathbf{v}_{\text{proj}}.
\end{equation}

\noindent \textbf{Introducing a risk-aware adaptive intervention mechanism}.
To prevent performance degradation on benign inputs from fixed-strength interventions, we introduce the Harm-Sensing Classifier. 
This component assesses the harmfulness of inputs in real time and dynamically adjusts the intervention strategy. 
The classifier adopts a Multi-Layer Perceptron architecture, and its input feature vector is constructed by concatenating the hidden states of the final token from the last $K$ layers of the model:
\begin{equation}
\label{eq:feature_vector}
\mathbf{z} = \text{Concat}(\mathbf{a}_{L-K+1}, \ldots, \mathbf{a}_{L}),
\end{equation}
where $\mathbf{a}_{i}$ denotes the activation of the final token in the $i$-th layer.
The classifier is trained using the standard Binary Cross-Entropy (BCE) loss, defined as:
\begin{equation}
\label{eq:bce_loss}
\mathcal{L}_{\text{BCE}}(y, p) = -[y \log(p) + (1-y) \log(1-p)],
\end{equation}
where $y \in {0,1}$ represents the ground-truth label, and $p$ is the predicted probability that the input is malicious.
To further address the imbalance between easy and difficult samples, we employ Focal Loss~\cite{lin2017focal}. This loss function modifies BCE by introducing a modulation factor that emphasizes hard examples:
\begin{equation}
\label{eq:focal_loss}
\text{FL}(p_t) = -\alpha_t (1 - p_t)^{\gamma} \log(p_t),
\end{equation}
where $p_t$ is the predicted probability for the correct class, $\gamma$ is a focusing parameter that reduces the weight of well-classified examples, and $\alpha_t$ is a balancing parameter between classes.

During inference, the classifier produces a risk score $p \in [0,1]$ indicating the likelihood that an input is malicious. Based on this score, we compute a bipolar intervention coefficient $\alpha$ as follows:
\begin{equation}
\label{eq:determinate_alpha}
\alpha = \alpha_{\text{init}} \cdot (2p - 1).
\end{equation}

This coefficient continuously adjusts the strength and direction of the intervention within the range $[-\alpha_{\text{init}}, \alpha_{\text{init}}]$. When $\alpha > 0$, the input is considered high-risk. In this case, the system applies a safety prefix to the model’s response and uses the refined steering vector $\mathbf{v}_{\text{final}}$ to steer the activation states toward safer directions.
On the other hand, when $\alpha < 0$, the input is treated as benign. The system disables the prefix and applies the steering vector in the opposite direction to enhance the informativeness and quality of the model’s response.

\begin{table*}[htbp]
    \centering
    \small
    \setlength{\tabcolsep}{6pt}
    \renewcommand{\arraystretch}{1.1}
    \definecolor{Gray}{gray}{0.95}
    \definecolor{HeaderGray}{gray}{0.85}
    \captionsetup{skip=5pt}
    
    \begin{tabular}{@{}cl*{5}{c}l!{\vrule width 0pt}@{}}
        \toprule
        \multirow{2}{*}[-1ex]{\textbf{Model}} & \multirow{2}{*}[-1ex]{\textbf{Method}} & \multicolumn{5}{c}{\textbf{Malicious scenarios}} & \multirow{2}{*}[-1ex]{\textbf{Avg. $\downarrow$}} \\
        \cmidrule(lr){3-7}
        & & \textbf{HateSpeech} & \textbf{Malware Generation} & \textbf{Physical Harm} & \textbf{Fraud} & \textbf{Pornography} & \\
         \midrule

        \multirow{5}{*}{{\textbf{LLaVA-v1.5}}} 
        & Vanilla      & 0.522 (47)          & 0.589 (53)              & 0.767 (69)              & 0.633 (57)          & 0.356 (32)          & 0.573 (258) \\
        & ASTRA        & \textbf{0.078 (7)}  & \underline{0.056 (5)}   & \underline{0.178 (16)}  & \underline{0.156 (14)} & 0.200 (18)            & \underline{0.133 (60)} \\
        & SCANS        & 0.322 (29)          & 0.222 (20)              & 0.467 (42)              & 0.378 (34)          & \underline{0.178 (16)} & 0.313 (141) \\
        & ETA          & \underline{0.100 (9)} & 0.200 (18)                & 0.378 (34)              & 0.222 (20)          & \underline{0.178 (16)} & 0.216 (97) \\
        & \cellcolor{Gray}\textbf{SafeSteer} & \cellcolor{Gray}\textbf{0.078 (7)} & \cellcolor{Gray}\textbf{0.022 (2)} & \cellcolor{Gray}\textbf{0.078 (7)} & \cellcolor{Gray}\textbf{0.044 (4)} & \cellcolor{Gray}\textbf{0.033 (3)} & \cellcolor{Gray}\textbf{0.051 (23)}  \\
        
        \midrule
        
        \multirow{5}{*}{{\textbf{MiniGPT-4}}} 
        & Vanilla      & 0.189 (17)          & 0.222 (20)              & 0.278 (25)              & 0.244 (22)          & 0.211 (19)          & 0.229 (103) \\
        & ASTRA        & 0.122 (11)          & 0.156 (14)              & 0.333 (30)              & 0.178 (16)          & 0.211 (19)          & 0.200 (90) \\
        & SCANS        & 0.089 (8)           & 0.067 (6)               & \underline{0.122 (11)}  & \textbf{0.044 (4)}  & \textbf{0.033 (3)}  & 0.071 (32) \\
        & ETA          & \textbf{0.000 (0)}    & \textbf{0.044 (4)}      & 0.133 (12)              & \underline{0.056 (5)} & \textbf{0.033 (3)}  & \underline{0.053 (24)} \\
        & \cellcolor{Gray}\textbf{SafeSteer} & \cellcolor{Gray}\underline{0.022 (2)} & \cellcolor{Gray}\underline{0.056 (5)} & \cellcolor{Gray}\textbf{0.078 (7)} & \cellcolor{Gray}\textbf{0.133 (4)} & \cellcolor{Gray}\underline{0.056 (5)} & \cellcolor{Gray}\textbf{0.051 (23)} \\
        
        
        
        \bottomrule
    \end{tabular}
    \caption{Evaluation of the ASR for different defense methods across the five malicious scenarios of MM-SafetyBench. Each scenario consists of 30 questions, and the ASR in each cell is the average result combining three input attack types: SD, TYPO, and SD-TYPO. In the table, the best (lowest ASR) and second-best results are marked in bold and with an underline, respectively. The values in parentheses indicate the number of successful jailbreak samples.}
     \label{tab2:mm_safetybench_asr}
\end{table*}



\subsection{Step 3: Intervention Injection During Generation}

In this step, the previously refined steering vector is applied in real time during the model's response generation process to guide the activation state toward a safer direction.

We select a target layer $l$ in the model as the intervention point and, during the generation of the first $n$ tokens, adjust the hidden state $h^l$ of that layer in real time based on the risk-aware coefficient $\alpha$. This process enables low-overhead, fine-grained control without modifying the model architecture.The specific update rule is as follows:
\begin{equation}
\label{eq:update_hidden_states}
h^{\text{l}} = h^{\text{l}} + \alpha \cdot \frac{\mathbf{v}^{\text{l}}_{\text{final}}}{\|\mathbf{v}^{\text{l}}_{\text{final}}\|} \cdot \|h^{\text{l}}\|.
\end{equation}

Here, $\mathbf{v}^{l}_{\text{final}}$ is the refined steering vector constructed in Step 2, which has been computed for the current input. This vector is normalized and injected into the current activation state proportionally, with its magnitude controlled by $\alpha$, thereby providing directional semantic steering during generation without altering the original network structure.

This operation is a simple vector addition with negligible computational cost and does not impact the model's inference efficiency.
As a result, SafeSteer significantly improves the suppression of high-risk inputs while preserving the model’s capabilities on benign instructions, achieving a balanced trade-off among safety, usability, and efficiency.

\section{Experiments}

\subsection{Experimental Setup}
\noindent\textbf{Datasets.}
We evaluate SafeSteer using malicious samples from MM-safetybench's~\cite{liu2024mm} Illegal-Activity category (40 samples) paired with blank images as original inputs, augmented through seven jailbreak attacks: Figstep~\cite{gong2025figstep}, Query-relevant~\cite{liu2024mm}, SI-Attack~\cite{zhao2025jailbreaking}, VAJM~\cite{qi2024visual}, Bap~\cite{ying2025jailbreak}, UMK~\cite{wang2024white}, and HADES~\cite{li2024images}. 
Details of the attack settings are provided in Appendix B.
Additional evaluation uses random 30 samples from five MM-safetybench malicious scenarios and MM-Vet~\cite{yu2023mm} benchmark to assess the model's benign performance.

\noindent\textbf{Evaluation Metrics.}
Our experimental evaluation centers on two dimensions: safety and model utility.
In terms of safety, we adopt the Attack Success Rate (ASR) as the primary evaluation metric. A ``successful jailbreak'' is adjudicated by GPT-4o.
Regarding model utility, we use scores on integrated capability benchmarks to measure the impact of the defense method on the model's performance.

\noindent\textbf{Baselines.}
We compare SafeSteer against SOTA baselines from two types of inference-time defense approaches.
For post-hoc evaluation pipelines, we selected ETA~\cite{ding2024eta}, which uses a two-stage framework of multimodal assessment and best-of-n search to correct unsafe responses.
For activation steering methods, we chose ASTRA~\cite{wang2025steering}, which builds vectors using image attribution to counter visual attacks, and SCANS~\cite{cao2025scans}, which employs conditional steering to defend against harmful queries while preserving model utility.
Details of the baseline settings are provided in Appendix C.

\noindent\textbf{Implementation Details.} 
Our experiments are primarily based on three open-source VLMs: LLaVA-v1.5-7B~\cite{liu2023visual}, MiniGPT4-7B (Llama 2)~\cite{zhu2023minigpt}, and Qwen2.5-VL-7B~\cite{bai2025qwen2}, using a single NVIDIA GeForce RTX 4090 GPU (24GB memory). 
Further hyperparameters are in the appendix A.

 \begin{table*}[htbp]
    \centering
    \small
    \setlength{\tabcolsep}{10pt}
    \renewcommand{\arraystretch}{1.1}
    \definecolor{Gray}{gray}{0.94}
    \definecolor{HeaderGray}{gray}{0.82}
    \definecolor{TaskHeaderGray}{gray}{0.88}
    \captionsetup{skip=5pt}
    
    \begin{tabular}{@{\extracolsep{\fill}}cl*{6}{c}c!{\vrule width 0pt}@{}}
        \toprule
        \multirow{2}{*}[-1ex]{\textbf{Model}} & \multirow{2}{*}[-1ex]{\textbf{Method}} & \multicolumn{6}{c}{\textbf{MM-Vet Capability Assessment}} & \multirow{2}{*}[-1ex]{\textbf{Avg. $\uparrow$}} \\
        \cmidrule(lr){3-8}
        & & \textbf{Recognition} & \textbf{OCR} & \textbf{Knowledge} & \textbf{Generation} & \textbf{Spatial} & \textbf{Math} &\\
        \midrule
        
        \multirow{5}{*}{\textbf{LLaVA-v1.5}}
        & Vanilla & \underline{32.8} & \underline{19.6} & 15.1 & 17.0 & \underline{21.7} & \textbf{11.5} & \underline{28.1} \\
        & ASTRA  & 29.7 & 15.7 & 12.9 & 12.4 & 18.4 & 7.7 & 25.3 \\
        & SCANS  & 29.3 & 15.0 & 10.4 & 9.6 & 21.1 & \textbf{11.5} & 24.6 \\
        & ETA    & 32.3 & 18.5 & \underline{15.4} & 16.9 & 21.3 & 7.7 & 27.3 \\
        & \cellcolor{Gray}\textbf{SafeSteer} & \cellcolor{Gray}\textbf{35.3} & \cellcolor{Gray}\textbf{20.0} & \cellcolor{Gray}\textbf{19.9} & \cellcolor{Gray}\textbf{21.6} & \cellcolor{Gray}\textbf{26.7} & \cellcolor{Gray}\underline{11.2} & \cellcolor{Gray}\textbf{29.8} \\
        
        \midrule
        
        \multirow{5}{*}{\textbf{MiniGPT-4}}
        & Vanilla & \underline{26.6} & \textbf{18.9} & \textbf{14.5} & \underline{14.4} & 20.5 & \textbf{14.6} & \textbf{23.3} \\
        & ASTRA  & 22.7 & 15.1 & 12.9 & 10.0 & \textbf{23.2} & 0.0 & 19.7 \\
        & SCANS  & 16.9 & 13.4 & 7.1 & 3.5 & 17.7 & 11.5 & 15.8 \\
        & ETA    & 26.2 & 15.8 & \underline{13.9} & \textbf{14.6} & 17.6 & \underline{12.7} & 22.4 \\
        & \cellcolor{Gray}\textbf{SafeSteer} & \cellcolor{Gray}\textbf{27.5} & \cellcolor{Gray}\underline{16.0} & \cellcolor{Gray}13.8 & \cellcolor{Gray}13.6 & \cellcolor{Gray}\underline{21.3} & \cellcolor{Gray}\underline{12.7} & \cellcolor{Gray}\underline{23.0} \\
        
        \midrule
        
        \multirow{5}{*}{\textbf{Qwen2.5-VL}}
        & Vanilla & \textbf{53.3} & \textbf{50.7} & \underline{48.8} & \textbf{49.2} & 46.1 & \textbf{53.5} & \textbf{52.6} \\
        & ASTRA  & 52.9 & 44.1 & \textbf{49.2} & 47.1 & 39.1 & 45.4 & 50.1 \\
        & SCANS  & 47.9 & 49.1 & 42.1 & 45.1 & \textbf{48.0} & 49.6 & 48.1 \\
        & ETA    & 51.5 & 49.8 & 46.7 & 45.5 & \underline{47.1} & \underline{53.1} & 51.9 \\
        & \cellcolor{Gray}\textbf{SafeSteer} & \cellcolor{Gray}\underline{53.1} & \cellcolor{Gray} \underline{50.3} & \cellcolor{Gray} 48 & \cellcolor{Gray} \underline{47.4} & \cellcolor{Gray}46.9 & \cellcolor{Gray} \underline{53.1} & \cellcolor{Gray} \underline{52.5}\\
        
        \bottomrule
    \end{tabular}
    \caption{Comparison of utility scores for different defense methods on the MM-Vet benchmark, reporting scores for six core vision-language capabilities. In the table, the best and second-best results are marked in bold and with an underline, respectively.}
    \label{tab3:utility_score}
\end{table*}

\begin{table}[!t]
    \centering
    \setlength{\tabcolsep}{4pt}
    \captionsetup{skip=5pt}
    
    \begin{tabular*}{\columnwidth}{@{\extracolsep{\fill}}l c c c}
        \toprule
        \multirow{2}{*}[-1ex]{\textbf{Method}} & \multirow{2}{*}[-1ex]{\textbf{Single Inference}} & \multicolumn{2}{c}{Inference Time (second)$\downarrow$} \\
        \cmidrule(lr){3-4} 
        & & \textbf{Original} & \textbf{MM-Vet} \\
        \midrule
        LLaVA-v1.5 & \checkmark & 4.26 & \textbf{1.71} \\
        ASTRA & \checkmark & 6.05 & 2.52 \\
        SCANS & \checkmark & 8.34 & 2.27 \\
        ETA & $\times$ & 20.45 & 4.27 \\
        SafeSteer & \checkmark & \textbf{2.42} & 2.05 \\
        \bottomrule
    \end{tabular*}
    \caption{Average inference time per response for different defense methods on LLaVA-v1.5 under two different scenarios. The Original dataset is used as the malicious scenario, and the MM-Vet dataset is used as the benign scenario.}
    \label{tab4:inference_time}
\end{table}

\subsection{Main Results}

\noindent\textbf{Analysis of Defense Performance and Generalization against Multiple Attacks.}
To rigorously stress-test SafeSteer, we built a challenging test set of eight mainstream attacks, covering diverse categories from black-box to white-box to represent the complex VLM threat landscape.
Against these varied attacks, SafeSteer demonstrated exceptional defense performance. The results in Table~\ref{tab1:jailbreak_asr} clearly shows that It achieved the lowest ASR levels across all tested models, with an average ASR of just 5.9\% on the vulnerable LLaVA-v1.5. 
This effectiveness stems from its mechanism of correcting harmful semantics directly within the activation space, rather than operating on input representations.
SafeSteer's defense generalization is powerfully quantified by its extremely low ASR standard deviation across diverse attacks.
As shown in the last column of Table~\ref{tab1:jailbreak_asr}, SafeSteer achieves the lowest standard deviation on all tested models, confirming its stable performance against different attack mechanisms.
This high stability validates that our method achieves true generalized defense by directly targeting common attack pathways within the activation space.

Furthermore, we evaluated SafeSteer on the five malicious scenarios within the MM-SafetyBench. 
The experimental results, as shown in Table~\ref{tab2:mm_safetybench_asr}, indicate that SafeSteer maintained robust defense capabilities across all scenarios, achieving a low total ASR of 5.1\%, which confirms its broad scenario coverage.
This result proves our ``safety subspace'' captures a generalizable representation of harmfulness rather than overfitting to specific scenarios.
In addition, we tested SafeSteer's effectiveness against adaptive attacks, with detailed results provided in the appendix E.

\noindent\textbf{Comparative Analysis with Existing VLM Defenses.} 
SafeSteer's superiority is evident not just in its performance but in how it resolves the inherent deficiencies of current VLM defenses.
One prevailing method, exemplified by ETA, uses a multi-step evaluation pipeline whose multiple interactions with an external evaluator inevitably cause high inference latency.
The data in Table~\ref{tab4:inference_time} clearly reflects this, showing SafeSteer's inference efficiency to be far superior to that of ETA. 
In terms of defense effectiveness (see Table~\ref{tab1:jailbreak_asr}),ETA is also significantly inferior, with an ASR of 20.0\% on vulnerable models like LLaVA-v1.5 compared to SafeSteer's 5.9\%.
This suggests that when the model's own defenses are unreliable, such post-hoc evaluation mechanisms are more prone to failure. 
Regarding model utility(see Table~\ref{tab3:utility_score}), while ETA's score surpasses other baselines, it still falls short of SafeSteer's almost lossless performance.

Another defense method is the activation engineering methods, which rely on a crude steering vector generated through simple averaging (\textit{e.g.} ASTRA and SCANS). 
In Table~\ref{tab1:jailbreak_asr}, demonstrating the superiority of our refined vector, SafeSteer achieves a substantially lower ASR than these methods.
Furthermore, in terms of model utility(see Table~\ref{tab3:utility_score}), both ASTRA and SCANS exhibit varying degrees of decline in their utility scores, indicating that these methods severely impair the model's performance while suppressing attacks.
Meanwhile, SafeSteer maintains superior inference efficiency over other activation steering methods across both benign and malicious tasks, as detailed in Table~\ref{tab4:inference_time}.
We also compare against a safety fine-tuning method, with details provided in the appendix D.

In summary, these results validate that SafeSteer achieves a synergistic improvement in efficiency, effectiveness, and utility for defending against VLM jailbreak attacks.

\begin{table}[!t]
    \centering
    \setlength{\tabcolsep}{5pt}
    
    \begin{tabular}{lcc}
    \toprule
    \textbf{Configuration} & \textbf{ASR} & \textbf{Utility Score} \\
    \midrule
    Vanilla Model           & 0.628(201) & 28.1 \\
    \textbf{+} Safety Prefix        & 0.212(68)  & 9.2  \\
    \textbf{+} Classifier            & 0.212(68)  & 27.8 \\
    \textbf{+} Raw Steering          & 0.134(43)  & 28.0 \\
    \textbf{+} Subspace Reconstruction   & 0.109(35)  & 29.7 \\
    SafeSteer(Ours)            & \textbf{0.059(19)}   & \textbf{29.8} \\
    \bottomrule
    \end{tabular}
    \captionsetup{skip=5pt}
    \caption{Ablation study of SafeSteer's components. This table shows the impact on ASR and Utility score as each component is added incrementally to a vanilla model. All experiments are conducted on LLaVA-v1.5.}
    \label{tab5:ablation}
\end{table}

\subsection{Ablation Studies}
\noindent\textbf{Component-wise Analysis.}
To dissect the contribution of each component within SafeSteer, we conducted a comprehensive ablation study on the LLava-v1.5 model. 
We incrementally added each component and evaluated its impact on the ASR and utility score in MM-Vet. 
The results, presented in Table~\ref{tab5:ablation}, demonstrate a clear and logical progression of performance improvement.

The study begins with the Vanilla Model, which serves as our baseline with a high ASR of 62.8\%. Applying a universal Safety Prefix to all inputs drastically reduces the ASR to 21.2\% but at the severe cost of collapsing the utility score to 9.2. 
The addition of the Classifier rectifies this by applying the prefix selectively, restoring utility to 27.8 while maintaining the strong defense.

Building on this adaptive foundation, incorporating raw steering vector further lowers the ASR to 13.4\%.
The effectiveness of our core technical innovation is validated in the next step, which is to decompose and reconstruct the raw steering vector using a secure subspace. This not only improves the ASR to 10.9\% but also boosts the utility score to 29.7, surpassing the vanilla model. Finally, the complete SafeSteer, which incorporates the linear adjustment of the strength coefficient $\alpha$, achieves the best result. 
It drives the ASR down to a remarkable 5.9\% while maintaining a high utility score of 29.8. 
These results conclusively demonstrate that each component of SafeSteer plays a crucial and synergistic role in achieving an exceptional balance between robust security and high utility.

\begin{figure}[!t]
    \centering
    \includegraphics[width=1.0\linewidth]{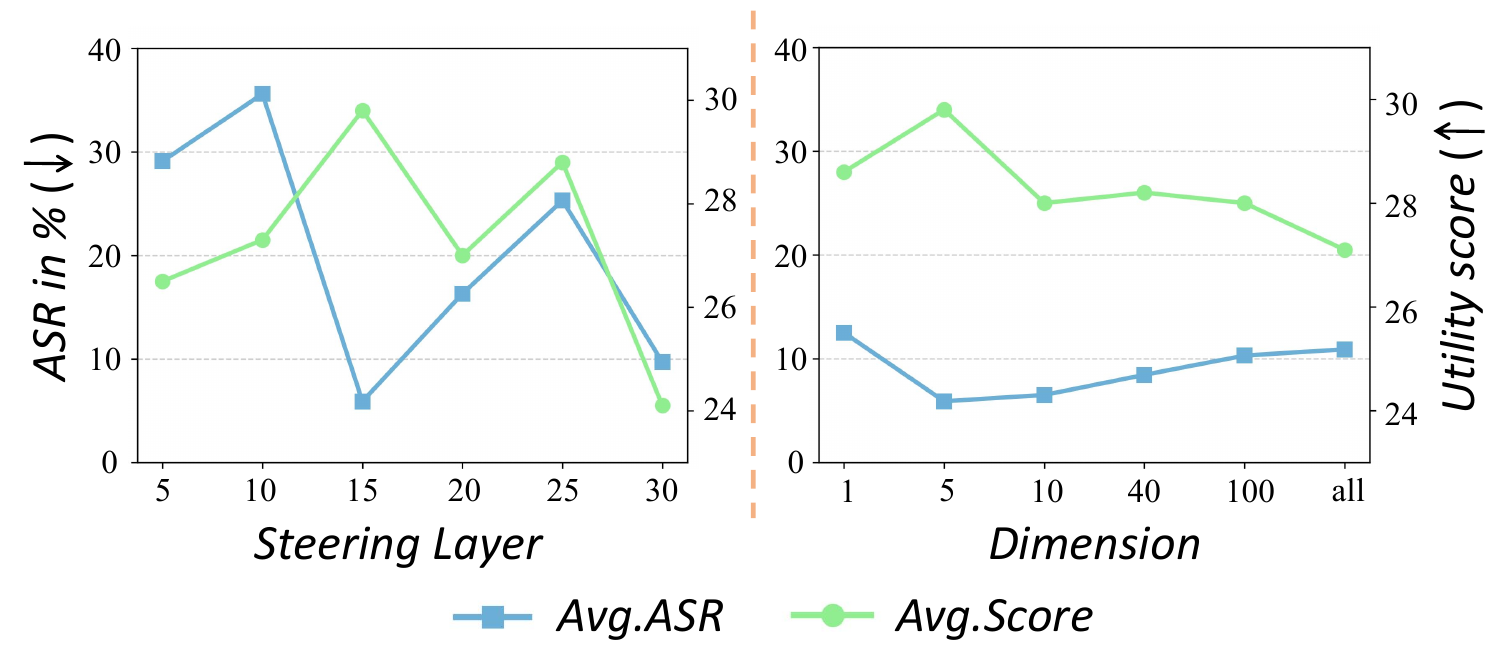}
    \captionsetup{skip=5pt}
    \caption{Impact of the steering layer (left) and subspace dimension (right) on LLaVa-v1.5, showing the trade-off between average ASR and Utility score.}
    \label{figure2}
\end{figure}

\begin{table}[!t]
    \centering
    \setlength{\tabcolsep}{3.5pt}
   
    \begin{tabular}{l c c c}
    \toprule
    \textbf{Model} & \textbf{JailbreakV-28k} & \textbf{Multi-Attack} & \textbf{Average} \\
    \midrule
    LLaVA-v1.5 & 0.997 & 0.996 & 0.997 \\
    MiniGPT-4 & 0.998 & 0.999 & 0.999 \\
    Qwen2.5-VL & 0.996 & 0.993 & 0.995 \\
    \bottomrule
    \end{tabular}
    
    \caption{AUROC performance of the harm-sensing classifier. The classifier is evaluated on its ability to distinguish benign inputs (from MM-Vet) from two malicious sets.}
    \label{tab:classifier_auroc}
\end{table}

\noindent\textbf{Validation of the Harm-Sensing Classifier.}
To validate the effectiveness of our proposed Harm-Sensing Classifier, we designed an experiment to evaluate its ability to distinguish malicious from benign inputs, using the Area Under the Receiver Operating Characteristic curve (AUROC) as the performance metric.
The malicious dataset comprised two components: a multi-source set with eight types of jailbreak attacks, and a randomly selected subset of 300 samples from the portion of the JailbreakV-28k dataset not used in training. 
In the evaluation, we used the MM-Vet dataset as the general benign set and paired it with each of the two malicious datasets to construct two test sets.
The results, presented in Table 6, show that our classifier obtained exceptionally high AUROC scores across all models, achieving an average of over 0.995.
This powerfully demonstrates the classifier's precision and reliability in identifying potential malicious inputs. 
Consequently, this provides a solid foundation for SafeSteer to implement adaptive intervention strategies, enabling robust defense against attacks while preserving the model's performance on benign tasks.

\noindent\textbf{Impact of Steering Layer Selection.}
To identify the optimal point for intervention within the model's architecture, We conducted an experiment to evaluate the effect of applying steering at different layers.
The results, presented in the left panel of Figure~\ref{figure2}, reveal a clear trend: the middle layers are the most effective for applying activation steering. 
Specifically, we observe that applying the steering near layer 15 yields the best performance, achieving the lowest ASR and the highest utility score simultaneously.
In contrast, steering at either the early layers or the final layers will affect the quality of model generation.

This finding aligns with recent research\cite{wang2025towards}. 
The middle layers are where the model is thought to form the core semantic concepts of its answer. 
By intervening at this critical layers, our steering mechanism can effectively guide the model’s conceptual direction towards safety, thus achieving the optimal balance between robust defense and preserved utility.
The early layers are primarily responsible for the fundamental fusion of visual and textual features.
An intervention at this layers can disrupt the model's basic comprehension of the input, thereby degrading performance on both safety and utility.
At the Late layers, the model is focused on superficial refinements like grammar and formatting.
While intervention here can corrupt the final output, it is largely unable to reverse a committed decision to generate harmful content. 

\begin{figure}[!t]
    \centering
    \includegraphics[width=1.0\linewidth]{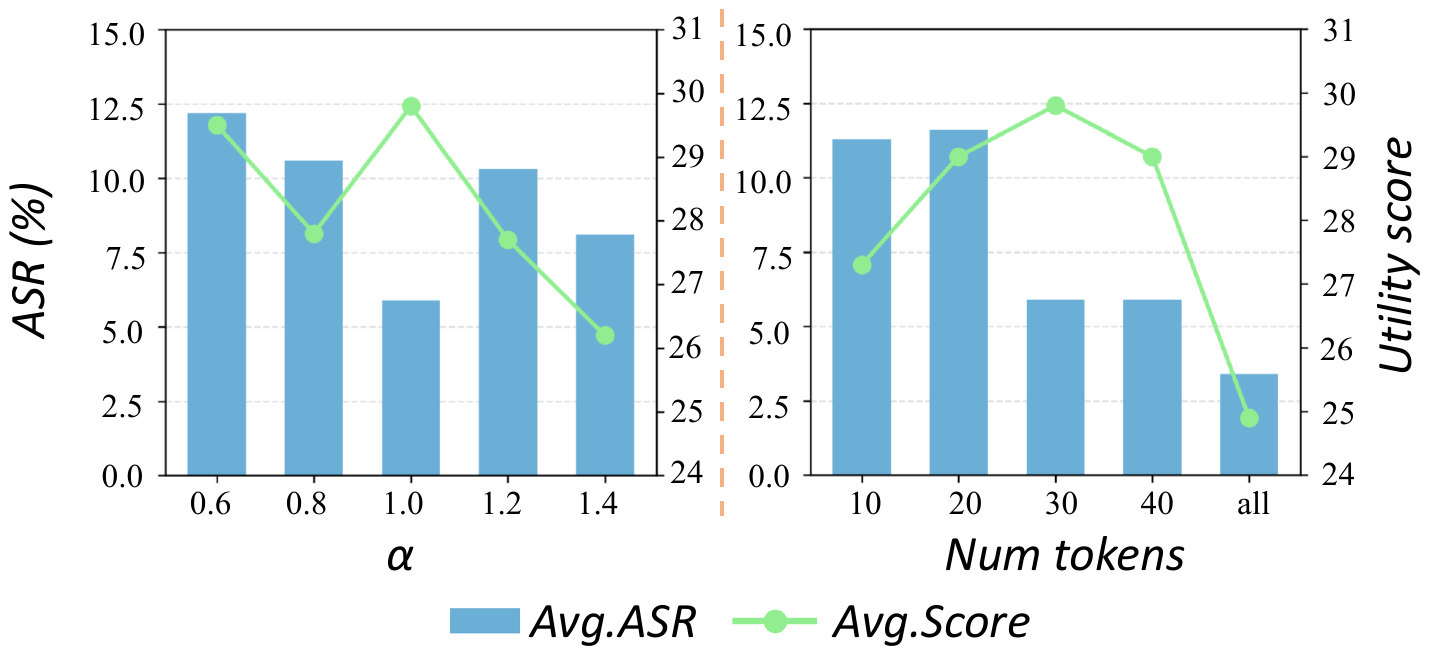}
    \captionsetup{skip=5pt}
    \caption{Performance of SafeSteer on the LLaVA model when varying the steering strength ($\alpha$, left) and the number of steered tokens (num, right). }
    \label{figure3}
\end{figure}

\noindent\textbf{Analysis of Subspace Dimension.}
The dimensionality of the safety subspace, D, is a critical parameter that reveals a clear signal-to-noise trade-off in the safety representation, as shown in the right panel of Figure~\ref{figure2}. 
The results demonstrate that a compact, low-dimensional subspace is optimal. Specifically, when the dimension is 5 achieves the best performance, yielding both the lowest ASR and the highest utility score. 
This finding provides a key insight: the first few principal components of the activation differences capture the most critical, high-signal safety concepts. 
Increasing the dimension beyond this point incorporates lower-variance components, which are more likely to be noise that contaminates the steering vector and degrades performance.

\noindent\textbf{Analysis of Steering Strength.}
The left panel of Figure~\ref{figure3} shows the performance trade-off as we vary $\alpha_{init}$ . 
The results indicate a clear optimal point. A value of $\alpha_{init}=1.0 $ achieves the best balance, yielding the lowest ASR while simultaneously maximizing the utility score. 
Values that are too low or too high result in a suboptimal defense, an unnecessary drop in utility, or both. 
This validates our choice of using $\alpha_{init}=1.0$ for the LLaVA-v1.5 model.

\noindent\textbf{Analysis of Intervention Depth.}
The right panel of Figure~\ref{figure3} illustrates the impact of the number of steered tokens, revealing steering more tokens generally improves defense but impacts the model's performance.
For the LLaVA model, our results indicate that steering the first 30 tokens is optimal, achieving near-optimal ASR while maximizing the utility score.
This suggests a response's core direction is determined early in the generation phase, and prolonged intervention yields diminishing safety returns while increasingly degrading performance on benign tasks.

\section{Conclusion}
This paper proposes SafeSteer, an efficient and effective adaptive defense framework, designed to address the core challenges in VLMs safety. 
We first reveal that the fundamental limitation of existing activation steering methods lies in their crude processing of steering vectors, and we innovatively introduce the ``safety subspace'' concept to achieve vector purification.
Building on this, SafeSteer employs a Sense-and-Intervene mechanism that combines precise steering vector with adaptive strength adjustment, thereby achieving a balance of security, utility, and efficiency. 
Extensive experiments demonstrate that this method significantly reduces the attack success rate while maintaining, and even enhancing, the model's performance with minimal additional overhead. 
We hope that our work on adaptive, precise steering will inspire future research into more accurate and reliable VLM safety methods.


\bibliography{aaai2026}
\end{document}